# The XML and Semantic Web Worlds: Technologies, Interoperability and Integration. A Survey of the State of the Art *

**Nikos Bikakis**[1] • **Chrisa Tsinaraki**[2] • **Nektarios Gioldasis**[2] •
**Ioannis Stavrakantonakis**[2] • **Stavros Christodoulakis**[2]

[1] National Technical University of Athens, Greece &
Institute for the Management of Information Systems,
"Athena" Research Center, Greece.
e-mail: *bikakis@dblab.ntua.gr*

[2] Lab. of Distributed Multimedia Information Systems & Applications,
Technical University of Crete, Greece.
e-mail: *{chrisa, nektarios, gstavrak, stavros}@ced.tuc.gr*

**Abstract**

In the context of the emergent *Web of Data*, a large number of organizations, institutes and companies (e.g., *DBpedia*, *Geonames*, *PubMed ACM*, *IEEE*, *NASA*, *BBC*) adopt the *Linked Data* practices and publish their data utilizing Semantic Web (SW) technologies. On the other hand, the dominant standard for information exchange in the Web today is XML. Many international standards (e.g., *Dublin Core*, *MPEG-7*, *METS*, *TEI*, *IEEE LOM*) have been expressed in XML Schema resulting to a large number of XML datasets. The SW and XML worlds and their developed infrastructures are based on different data models, semantics and query languages. Thus, it is crucial to provide interoperability and integration mechanisms to bridge the gap between the SW and XML worlds.

In this chapter, we give an overview and a comparison of the technologies and the standards adopted by the XML and SW worlds. In addition, we outline the latest efforts from the W3C groups, including the latest working drafts and recommendations (e.g., OWL 2, SPARQL 1.1, XML Schema 1.1). Moreover, we present a survey of the research approaches which aim to provide interoperability and integration between the XML and SW worlds. Finally, we present the Sparql2XQuery and Xs2Owl Frameworks, which bridge the gap and create an interoperable environment between the two worlds. These Frameworks provide mechanisms for: (a) Query translation (*SPARQL to XQuery translation*); (b) Mapping specification and generation (*Ontology to XML Schema mapping*); and (c) Schema transformation (*XML Schema to OWL transformation*).

## 1. Introduction

In the context of the emerging *Web of Data*, a large number of organizations, institutes and companies (e.g., *DBpedia*, *Geonames*, *PubMed*, *DBLP*, *ACM*, *IEEE*, *IBM*, *NASA*, *BBC*, *NSF*, etc.) adopt the *Linked Data* practices; they use the Semantic Web (SW) technologies, publish their data and offer SPARQL endpoints (i.e., SPARQL-based search services) over it. On the other hand, the dominant standard for information exchange in the Web today is XML. In addition, many international standards (e.g., *MPEG-7*, *MPEG-21*, *VRA Core*, *METS*, *TEI*, *IEEE LOM*, etc.) have been expressed in XML Schema, resulting to a large number of XML datasets.

In the Web of Data, the applications and services have to coexist and interoperate with the existing applications that access legacy systems like, for instance, the very large XML-based audiovisual digital libraries of the digital video broadcasters (e.g., *BBC TV-Anytime Service*, etc.), the XML-based digital libraries of the cultural heritage institutions, etc.

Since the SW (*OWL/RDF/SPARQL*) and XML (*XML Schema/XML/XQuery*) worlds have different data models, different semantics and use different query languages to access them, it is crucial to develop tools and methodologies that will allow bridging the gap between them. In addition, it is unrealistic to expect that all the legacy data (e.g., Relational, XML, etc.) will be converted to SW data. Thus, it is crucial to provide interoperability and integration mechanisms that will allow the SW users to access external heterogeneous data sources from their own working environment. It is also important to offer SPARQL endpoints over legacy data in the Linked Data context.

In this chapter, we deal with the mechanisms that allow the exploitation of the legacy data, in the Web of Data. In particular, in the first part of the chapter (Section 2), we try to present and compare the XML and SW worlds. In particular, we outline and compare the technologies and the standards adopted in the two worlds. In addition, we present the latest efforts from the W3C groups, including the latest working drafts and recommendations (e.g., XML Schema 1.1, OWL 2, SPARQL 1.1, XQuery and XPath Full Text 1.0, XQuery Update Facility 1.0, etc.).

In the second part (Sections 3 and 4), we present a survey of the existing approaches that deal with the interoperability and integration issues between the XML and SW worlds.

Finally, in the third part (Sections 5–8), we outline the SPARQL2XQuery and XS2OWL Frameworks that have been developed in order to provide an interoperable environment between the SW and the XML worlds.

## 2. XML World vs. Semantic Web World — A Comparison

In this section we outline the XML and SW worlds, we present the adopted technologies and we compare their basic characteristics. Throughout this comparison, we distinguish three levels: (a) The *data level*; (b) The *schema level*; and (c) The *query level*. Table 1 provides an overview of the current W3C standards adopted in each level in the XML and SW worlds. In addition Fig. 1 outlines the history of the XML and SW technologies proposed by W3C.

**Table 1. Overview of the W3C Standards currently adopted in the XML and SW Worlds**

| Level | XML World | Semantic Web |
|-------|-----------|--------------|
| **Data** | XML | RDF |
| **Schema** | XML Schema | RDF Schema – OWL |
| **Query** | XQuery – XPath | SPARQL |

At the data level, the *Extensible Markup Language* (*XML*) [1] is the data representation language in the XML world, while the *Resource Description Framework* (*RDF*) [2] is used to represent the SW data.

At the schema level, the *XML Schema* [3] is used to describe the structure of the XML data. Currently, the XML Schema 1.0 is a W3C recommendation and the XML Schema 1.1 [4][5] is under development and has reached the W3C working draft level of standardization. Regarding the SW, the *RDF Schema* (*RDFS*) [6] and the *Web Ontology Language* (*OWL*) [7] are exploited to describe the structure and the semantics of the RDF data. Recently, a new version of OWL, OWL 2.0 [8] has become a W3C recommendation.

Finally, at the query level, the *XML Path Language* (*XPath*) [9] and the *XML Query Language* (*XQuery*) [10] are employed for querying XML data. In the SW world, the *SPARQL Protocol and RDF Query Language* (*SPARQL*) [12] is the standard query language for RDF data. Currently, the W3C SPARQL working group [14] is working on the extension of the SPARQL query language in several aspects, resulting in SPARQL 1.1. SPARQL 1.1 includes several components like the *SPARQL 1.1 Query*, *Update*, *Protocol*, *Service Description*, *Property Paths*, *Entailment Regimes*, *Uniform HTTP Protocol for Managing RDF Graphs*, and *Federation Extensions*.

In the rest of this section, we outline the models adopted in the XML and SW worlds at the data level (Section 2.1) and we present and compare their schema (Section 2.2) and query (Section 2.3) level languages.

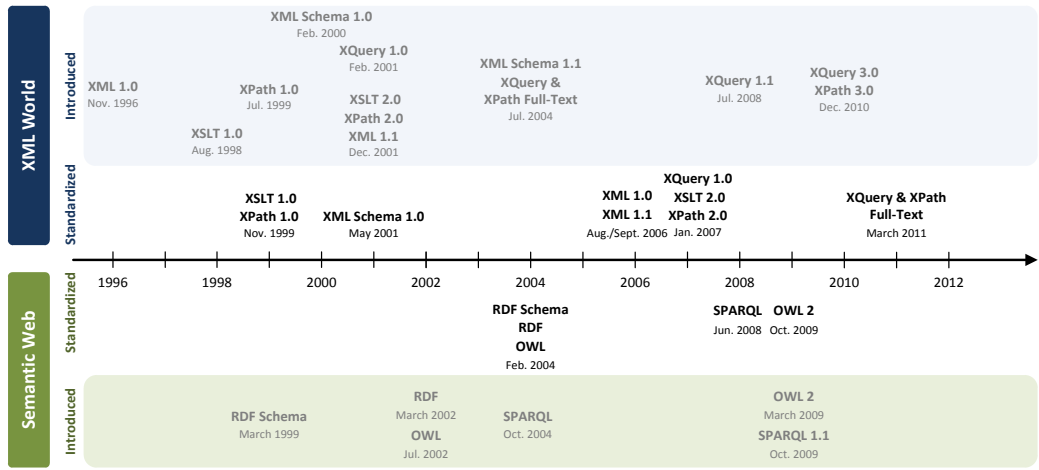

**Fig. 1.** Timeline of the W3C technologies related to the XML and SW Worlds [1]

## 2.1. Data Level

In this section we present in brief the models adopted at the data level in the XML and SW worlds. Section 2.1.1 describes the XML model that is adopted in the XML world, while Section 2.1.2 outlines the RDF model that is used in the SW world.

### 2.1.1. XML

The *Extensible Markup Language* (*XML*) [1] is a general-purpose markup language, designed to describe structured documents. XML is based on tags like HTML, however, XML it does not have a fixed set of tags, but allows users to define their own tags. In addition, unlike HTML, the XML tags have no specific semantics. An XML document consists of plain text and markup, in the form of tags and may be represented as an ordered labeled tree. An XML document may contain the following types of nodes: document nodes, elements, attributes, text nodes, comments, processing instructions, and namespaces.

### 2.1.2. RDF

The *Resource Description Framework* (*RDF*) [2] is a general purpose language for representing information about resources on the Web. To some extent, RDF is a lightweight ontology language. RDF has a very simple and flexible data model, based on the central concept of the RDF statement. RDF statements are triples

---

(subject, predicate, object) consisting of the resource (the subject) being described, a property (the predicate), and a property value (the object). In particular, the subject can either be an IRI or a Blank node. Every predicate must be an IRI and the object can be an IRI, Blank node or RDF Literal. A collection of RDF statements (or else RDF triples) can be intuitively understood as a directed labeled graph, where the resources are nodes and the statements are arcs (from the subject node to the object node) connecting the nodes. Finally a set of RDF triples is called RDF Dataset or RDF Graph.

## 2.2.    Schema Level

In this section we present the schema languages adopted in the XML and SW worlds. Section 2.2.1 outlines the XML Schema language, used in the XML world. The SW languages are then presented; in particular, RDF Schema is presented in Section 2.2.2 and OWL language is presented in Section 2.2.3. Finally, a comparison between the schema languages adopted by the XML SW worlds is presented in Section 2.2.4.

### 2.2.1.    XML Schema

The *XML Schema* [3] is a schema definition language that has been developed by the W3C and has been expressed in XML syntax. XML Schema is intended to describe the structure and constrain the content of documents written in XML by providing rich structuring capabilities. XML Schema can specify the exact element hierarchy and specify various types of constraints placed on the XML data define options (e.g., limits on the number of occurrences of an element, ranges of the values of elements or attributes, etc.).

In particular, an XML document is composed of elements, with the root element delimiting the beginning and the end of the document. Each XML Schema element belongs to an XML Schema type, specified in the type attribute. The elements are distinguished into complex and simple elements, depending on the kind (simple or complex) of the types they belong to. The XML Schema simple types are usually defined as restrictions of the basic datatypes provided by XML Schema (i.e., xs:string, xs:integer, etc.). Moreover, the XML Schema complex types represent classes of XML constructs that have common features, represented by their elements and attributes.

Regarding the document structure, XML Schema, support rich structuring capabilities. The XML Schema elements may either have a predefined order as is specified in the XML Schema element sequence or be unordered as is specified in the XML Schema elements choice and all. The XML Schema sequence, choice and all elements may be nested. The minimum and maximum number of occurrences of the elements, choices and sequences are specified, respectively, in the minOccurs and maxOccurs attributes (absent minOccurs and/or maxOccurs correspond to values of 1). Reusable complex structures, combining sequence, choice and all elements may be defined as model groups. Finally, the reuse of element definitions is supported by the substitutionGroup attribute, which states that the current element is a specialization of another element.

The XML Schema attributes describe element features with values of simple type and may form attribute groups comprised of attributes that should be used simultaneously.

Default and fixed values may be specified for both attributes and simple type elements, in the default and fixed attributes respectively. Inheritance is supported for both simple and complex types, and the base types are referenced in the base attribute of the type definitions. All the XML Schema constructs may have textual annotations, specified in their annotation element.

The top-level XML Schema constructs (attributes, elements, simple and complex types, attribute and model groups) have unique names (specified in their name attribute). The nested elements and attributes have unique names in the context of the complex types in which they are defined, while the nested (complex and simple) types are unnamed. All the XML Schema constructs may have unique identifiers (specified in their id attribute). The top-level constructs may be referenced by other constructs using the ref attribute.

**XML Schema 1.1.** Currently, a new version of XML Schema, XML Schema 1.1 [4][5] is under development and has reached the W3C working draft level of standardization. XML Schema 1.1 is backwards compatible with XML Schema 1.0 and introduces several new features and mechanisms. The most important among them are discussed in the following paragraphs.

The new assert element is used to make assertions about element and attribute values, specify relationships and enforce constraints on elements and/or attributes above and beyond the constraints specified in their declarations.

The alternative element allows an element to have as type a member of a set of types.

The override element replaces, in XML Schema 1.1, the deprecated XML Schema 1.0 redefine element.

The all element has been enhanced in XML Schema 1.1 to allow elements with multiple occurrences. The substitutionGroup element has also been modified in XML Schema 1.1 and allows an element to be substituted by multiple elements. Finally, the any and anyAttribute elements have been enriched with additional attributes that allow to indicate the extension elements or attributes not allowed in an element.

Regarding the datatypes, XML Schema 1.1 introduces several new features: (a) The new error datatype is used to trigger an error; (b) The anyAtomicType is introduced, which represents the union of the value spaces of all the primitive types; (c) The XML Schema 1.1 dateTimeStamp datatype is introduced, which is identical to the dateTime, except from that it requires the time zone to be specified; and (d) The XML Schema 1.1 yearMonthDuration and dayTimeDuration datatypes are introduced, which are constrained versions of the duration datatype, and several others.

Finally, several facets are introduced, like, for instance, the assertion facet, which is used to constrain a simpleType; the explicitTimezone facet which is used with date datatypes to specify whether the time zone is required; and the minScale and maxScale facets which are used with the XML Schema 1.1 precisionDecimal datatype, in order to constrain the size of the exponent.

### 2.2.2. RDF Schema

The *RDF Schema* (*RDFS*) [3] is an extension of RDF designed to describe, using a set of reserved words called the RDFS vocabulary, resources and/or relationships between resources. It provides constructs for the description of types of objects (classes), type hierarchies (subclasses), properties that represent object features (properties) and property hierarchies (subproperty).

In particular, a Class in RDFS corresponds to the generic concept of a type or category, somewhat like the notion of a class in object-oriented languages, and is defined using the construct rdfs:Class. The resources that belong to a class are called its instances. An instance of a class is a resource having an rdf:type property whose value is the specific class. Moreover, a resource may be an instance of more than one classes. Classes can be organized in a hierarchical fashion using the construct rdfs:subClassOf. A property in RDFS is used to characterize a class or a set of classes and is defined using the construct rdf:Property. The rdfs:domain construct is used to indicate that a particular property applies to a designated class, and the rdfs:range construct is used to indicate the values of a particular property. In a similar way with the class hierarchies, RDFS provides rdfs:subPropertyOf construct for the definition of property hierarchies.

### 2.2.3. OWL

The *Web Ontology Language* (*OWL*) [7][8] is the standard language for defining and instantiating Web ontologies. OWL and RDFS have several similarities. OWL is defined as a vocabulary like RDF, however OWL has richer semantics.

An OWL Class is defined using the construct owl:Class and represents a set of individuals with common properties. All the OWL classes are considered to be subclasses of the class owl:Thing and superclasses of the class owl:Nothing. Moreover, OWL provides additional constructors for class definition, including the basic set operations, union, intersection and complement that are implemented, respectively, by the constructs owl:unionOf, owl:intersectionOf, and owl:complementOf and several other constructors like, for example, owl:oneOf, owl:equivalentClass, etc. Regarding the individuals, OWL allows to specify two individuals to be identical or different through the owl:sameAs and owl:differentFrom constructs. Unlike RDF Schema, OWL distinguishes a property whose range is a datatype value from a property whose range is a set of resources. Thus, the OWL Datatype properties are relations between class individuals and XML schema datatypes and are defined using the construct owl:DatatypeProperty; The OWL Object properties are relations between class individuals and are defined using the owl:ObjectProperty construct.

Finally, two properties may be stated to be equivalent, using the construct owl:equivalentProperty.

**OWL 2.** The *OWL 2 Web Ontology Language* (*OWL 2*) [8] is a W3C recommendation since the October of 2009. OWL 2 has a very similar overall structure with OWL 1 and is backwards compatible with it, while it introduces a plethora of new features.

Some of these new features are commonly referred as *syntactic sugar*, since these features do not change the expressiveness or the semantics of OWL 1; they have been introduced in order to make some common statements easier to be constructed like, for instance, the owl:AllDisjointClasses and owl:disjointUnionOf that are used to specify the disjoint classes more easily than using the owl:disjointWith and owl:unionOf OWL 1 statements.

In addition, a large number of the new OWL 2 features enhance the language expressiveness by adding restrictions and new characteristics over the OWL properties. Among them there are the qualified property cardinality restrictions that are specified using the new constructs owl:minQualifiedCardinality, owl:maxQualifiedCardinality and owl:qualifiedCardinality. Moreover, reflexive, irreflexive, and asymmetric properties are supported using, respectively, the owl:ReflexiveProperty, owl:IrreflexiveProperty and owl:AsymmetricProperty constructs. In addition, a set of classes can be specified to be mutually disjoint using the owl:propertyDisjointWith and owl:AllDisjointProperties constructs. Furthermore, the new property owl:hasSelf has been introduced to allow relating a class to itself. OWL 2 also allows stating that an individual should not hold certain values for certain properties using the new owl:NegativeObjectPropertyAssertion and owl:NegativeDataPropertyAssertion statements. Finally, a very useful, new feature allows a property to be defined in terms of a chain of object properties, using the construct owl:propertyChainAxiom.

The newly introduced owl:hasKey construct can be used to define that the instances of a class is identified by a set of datatype or object properties, thus providing an identity constraint mechanism.

OWL 1 is highly dependent on XML Schema regarding both the built-in and the user-defined datatypes. Using OWL 2, the users can integrate in their ontologies the datatype definitions, using the XML Schema datatypes and constraints. Moreover, OWL 2 has introduced two new built-in types: owl:real and owl:rational. Finally, OWL 2 allows new datatypes to be defined as the complement of existing datatypes, using the owl:datatypeComplementOf statement.

OWL 2 provides top and bottom object and datatype properties analogous to the owl:Thing and owl:Nothing classes for representing the universal and the empty class. In particular, OWL 2 provides the owl:topObjectProperty and owl:bottomObjectProperty properties, corresponding to the universal and the empty object property respectively, as well as the owl:topDataProperty and owl:bottomDataProperty properties, corresponding to the universal and the empty datatype property respectively.

### 2.2.4. Schema Level Comparison

In this section we provide a brief comparison between the schema definition languages adopted by the XML the SW worlds. Our comparison is based on a set of generic characteristics over the schema languages presented in the previous sections.

Table 2 presents an overview of this comparison. For each language we present: (a) Its Model Type; (b) Its Concrete Syntax, that is, how language elements are represented; (c) The Basic Constructs of its model; (d) The language Semantics; (e) The Identity Constraint mechanism supported by the schema definition languages; and (f) An overview of the User-Defined Datatypes mechanism (if such a mechanism is supported).

**Table 2. Comparison of the Schema Definition Languages in terms of a set of generic Characteristics**

| Characteristic | Schema Definition Languages | | | |
| --- | --- | --- | --- | --- |
| | **XML Schema** | **RDF/S** | **OWL 1** | **OWL 2** |
| **Model Type** | Hierarchical | Direct Label Graph | Direct Label Graph | Direct Label Graph |
| **Concrete Syntax** | XML | RDF/XML, N3, N-Triples, Turtle | RDF/XML, N3, N-Triples, Turtle | OWL/XML, Functional, Manchester, RDF/XML, N3, N-Triples, Turtle |
| **Basic Constructs** | Simple type, Complex type, Attribute, Element, Attribute group, Sequence Choice Annotation Extension, Restriction, Unique, Key, Keyref, Substitution Group, *Alternative*[+], *Assert*[+], *Override*[+], *Redefine*[+], etc. | Statement, Class, Property, Resource, type, subject, predicate, object, subClassOf, subPropertyOf, domain, range, Datatype, Literal, Bag, Seq, List, Alt, BlankNode | Class, Datatype Property, Object Property, Individual, Thing, Nothing, equivalentClass, intersectionOf, unionOf, complementOf, disjointWith, minCardinality, sameAs, oneof, hasValue, TransitiveProperty, FunctionalProperty, etc | InverseOf, NegativePropertyAssert ion, propertyChainAxiom, minQualifiedCardinality, hasSelf, AsymmetricProperty, AllDisjointClasses, disjoint, ReflexiveProperty, maxQualifiedCardinalit y, UnionOf, etc. |
| **Semantics** | Informal | Model Theory | Model Theory, RDF Graphs | Model Theory, Extension of *SROIQ DL* |
| **Identity Constraints** | Unique, Key, Keyref | — | — | hasKey |
| **User-Defined Datatypes** | minInclusive, minExclusive, maxLength, length, totalDigits, etc. | — | — | xsd:minInclusive, xsd:minExclusive, owl:onDatatype, owl:withRestrictions, etc. |

**Note.** The [+] indicates XML Schema 1.1 constructs.

It can be observed from Table 2 that the XML Schema provides a flexible mechanism to support identity constraints as well as rich capabilities for defining user datatypes. In contrast, none of the RDF Schema and OWL 1 languages supports identity constraints or user-defined datatypes. These limitations have been overcome by OWL 2, which introduces an identity constraint mechanism using the hasKey construct and supports user defined datatypes.

## 2.3.    *Query Level*

In this section we present the query languages used in the XML and SW worlds. Section 2.3.1 outlines the XPath, XQuery 1.0 and XQuery 1.1 languages, used in the XML world. The SPARQL query language that is adopted in the SW world is presented in Section 2.3.2. Finally, a comparison between the query languages adopted by the XML and SW worlds is presented in Section 2.3.2.

### 2.3.1.    XPath and XQuery

Path expressions play a significant role in XML query evaluation, since the path expressions are used to traverse the tree representations of the XML documents and select a node or a set of nodes. The *XML Path Language* (*XPath*) [9] is a W3C recommendation that specifies a path language capable of describing path expressions on the tree data model of XML. The XPath language is essentially a subset of the XQuery language, which is the W3C standard for querying XML data; thus, in the rest of this chapter we will only refer to XQuery language.

The *XML Query Language* (*XQuery*) [10] is based on a tree-structured model of the XML document content. XQuery exploits XPath expressions to address specific parts of the XML documents. The basic structure of the most XQuery queries is the FLWOR expression. FLWOR stands for the For, Let, Where, Order By and Return XQuery clauses. FLWORs, unlike path expressions, allow the users to manipulate, transform, and sort the query results.

The For and Let clauses generate a list of tuples preserving document order, with each tuple consisting of one or more bound variables. In particular, the For clause sets up an iteration over the tuples in the tuple list. The Let clause is used to set the value of a variable. However, unlike the For clause, it does not set up an iteration. The optional Where clause serves as a filter for the tuples generated by the For and Let clauses. The optional Order By clause is used to order the results; if no Order by clause exists, the order of the tuple list is determined by the For and Let clauses and by the document order. Finally, every XQuery expression has a Return clause that always comes last. The Return clause specifies the XML nodes that are included in the results and probably how they are formatted. In addition, XQuery supports conditional expressions based on the keywords if - else if - else.

A key feature of XQuery is the large number of built-in functions and operators provided (over 100) [11], covering a broad range of functionality That includes functions for manipulating strings, dates, combine sequences of elements, etc. Moreover, XQuery supports user-defined functions, defined either in the query itself, or in an external library. Both built-in and user-defined functions can be called from almost any place in a query.

### 2.3.2. SPARQL

The *SPARQL Protocol and RDF Query Language* (*SPARQL*) [15] is a W3C recommendation and it is today the standard query language for RDF data. The evaluation of SPARQL queries is based on graph pattern matching. A *Graph Pattern (GP)* is defined recursively and contains *Triple patterns* and SPARQL operators. The operators of the SPARQL algebra which can be applied on *graph patterns* are: AND (i.e., conjunction), UNION (i.e., disjunction), OPTIONAL (i.e., optional patterns, like left outer join) and FILTER (i.e., restriction). The *Triple patterns* are just like *RDF triples* except that each of the subject, predicate and object parts may be a variable. A sequence of conjunctive triple patterns is called *Basic Graph Pattern (BGP)*. The SPARQL Where clause consists of a *Graph Pattern* (*GP*).

SPARQL allows four query forms: Select, Ask, Construct and Describe. The Select query form returns a solution sequence, i.e., a sequence of variables and their bindings. The Ask query form returns a Boolean value (yes or no), indicating whether a query pattern matches or not. The Construct query form returns an RDF graph structured according to the *graph template* of the query. Finally, the Describe query form returns an RDF graph which provides a "description" of the matching resources. Thus, based on the query forms, the SPARQL query results may be *RDF Graphs*, *SPARQL solution sequences* and *Boolean* values.

SPARQL provides various solution sequence modifiers that can be applied on the initial solution sequence in order to create another, user desired, sequence. The supported SPARQL solutions sequence modifiers are: Distinct, Reduced, Limit, Offset and Order By. While the Distinct modifier ensures that duplicate solutions are eliminated from the solution set, the Reduced modifier simply allows them to be reduced. The Limit modifier puts an upper bound on the number of solutions returned. Moreover, the Offset modifier returns the solutions starting after the specified number of solutions. Finally, the Order By modifier, establishes the order of a solution sequence.

**SPARQL 1.1.** The SPARQL 1.1 is the result of the W3C SPARQL working group [14] on the extension of the SPARQL query language in several aspects. SPARQL 1.1 includes the components *SPARQL 1.1 Query*, *Update*, *Protocol*, *Service Description*, *Property Paths*, *Entailment Regimes*, *Uniform HTTP Protocol for Managing RDF Graphs*, and *Federation Extensions*.

The SPARQL 1.1 tries to eliminate the main limitations of the current SPARQL version like aggregate functions, nested queries, update operations, paths, and several other issues.

The aggregate functions that are supported from almost all the query languages will also be included in the SPARQL 1.1 Query [16]. In particular, SPARQL 1.1 will support the following aggregate functions: COUNT, SUM, MIN/MAX, AVG, GROUP_CONCAT, and SAMPLE.

In addition, nested queries, which are very important in cases where the result from one query is used as an input to another query, will be supported by SPARQL 1.1. Moreover, in order to implement negation, SPARQL 1.1 has adopted the new operators NOT EXISTS and MINUS.

In the current SPARQL version, a SELECT query may project only variables. SPARQL 1.1 allows the SELECT queries to project any SPARQL expression: Using the keyword AS in the SELECT clause, the result of a SPARQL expression is bound to the new variable specified by AS and this new variable is going to be projected.

In several cases, in order to find a node, it is necessary to express queries that use fixed-length paths to traverse the RDF graph. The SPARQL 1.1 Property Paths [18] extends the current basic graph patterns in order to support the expression of path patterns.

SPARQL 1.0 can be used only as a retrieval query language, since it does not support data management operators. The SPARQL Update 1.1 [17] includes several features for graph management. The INSERT and INSERT DATA operations are exploited to insert new triples in RDF graphs. In addition, the DELETE and DELETE DATA operations are used to delete triples from an RDF graph. The LOAD and CLEAR operations perform a group of update operations as a single action. In particular, the LOAD operation copies all the triples of a graph into the target graph, while the CLEAR operation deletes all the statements from the specified graph. Finally, in order to create a new RDF graph in an RDF graphs repository or to delete an RDF graph from an RDF graphs repository the CREATE and DROP operations are introduced.

### 2.3.3. Query Level Comparison

In this section we present a comparison between the query languages adopted in the XML and SW worlds. Our comparison has been based on a set of generic characteristics (Table 3) and on a set of language features (Table 4).

In Table 3 we give an overview of our comparison over a set of generic characteristics. For each query language, we present: (a) Its Language Type; (b) Its Input Data Model, that is, the data format accessed by the language; (c) The Basic Elements of the language; (d) The method exploited by the query language to evaluate the queries (Evaluation Method); (e) The Evaluation Clause, that is, the clauses of the query language that are used to describe the evaluation settings; (f) The algebra operators defined at each language (Evaluation Operators); The language Semantics; (g) The Query Types supported by the language; (h) The Result Form Declaration, that is, how and where the user can specify the form of the results inside a query; (i) The available result forms for each language (Results Form); (k) The Results Modification, that is, how the return results can be modified; (l) The mechanisms provided for defining evaluation Restrictions; (m) The built-in Operations & Functions provided by each query language; and (n) The Logic adopted by the language operations and functions.

It can be observed from Table 3 that a large number of differences can be identified. These differences are mainly due to: (a) The different data models types (i.e., tree – graph); (b) The functional style of the XQuery language; (c) The different assumptions made (i.e., the *Closed World Assumption* in the XML world vs. the *Open World Assumption* in the SW world); (d) The logic adopted (XQuery adopts Boolean logic while SPARQL adopts three-valued logic), and several others.

**Table 3. Comparison of the Query Languages in terms of a set of generic Characteristics**

| Characteristic | Query Languages | |
|---|---|---|
| | **XQuery** | **SPARQL** |
| **Language Type** | Functional, Declarative | Declarative |
| **Input Data Model** | XML | RDF |
| **Basic Elements** | XPaths expressions, For, Let, Where, Return , Order by, If - Else If – Else | Select, Construct, Ask, Describe, Union, Optional, Filter, Limit, Offset, Reduced, Distinct, Order by |
| **Evaluation Method** | Tree Traverse, Tree Matching | Graph Matching |
| **Evaluation Clause** | Everywhere in the Query | Where Clause |
| **Evaluation Operators** | No Standards Operators. Operators can be Implemented Using "Simple" Operators and Clauses (e.g., =, *If-Else If*, etc.) | Union, And, Optional, Filter |
| **Semantics** | Closed World Assumption (*CWA*) | Open World Assumption (*OWA*) |
| **Query Types** | Not Standard - Combination of FLOWR Expressions | Select, Construct, Ask, Describe |
| **Result Form Declaration** | Return Clause Syntax | Query Forms (i.e., Select, Ask, etc.) |
| **Result Form** | Flexible | Sequence of Variables Bindings, RDF Graph, Boolean |
| **Result Modifications** | Functions | Solution Sequence Modifiers |
| **Restrictions** | XPath Predicates, Where Clause Conditions, *If-Else If* | Triple patterns with constant parts, Filter |
| **Operations & Functions** | &&, ‖, !, =, !=, >, <, >=, <=, +, -, *, /, abs, floor, concat, substring, string-length, lower-case, starts-with, matches, replace, exists, distinct-values, insert-before, empty, count, avg, max, min, sum, etc. | &&, ‖, !, =, !=, >, <, >=, <=, +, -, *, /, bound, lang, regex, isIRI, isBlank, isLiteral, str, datatype, sameTerm, langMatches |
| **Logic** | Boolean Logic (True / False) | Three-valued Logic (True / False / Error) |

Table 4 presents a comparison between the XQuery, the SPARQL 1.0 and the SPARQL 1.1 query languages over several features. For each feature we indicate if the languages fully support it (✓), partially support it (❖) or do not support it (✗). Table 4 is based on the W3C specifications and working drafts and does not consider possible languages extension proposed from other parties.

It can be observed from Table 4 that the XQuery language supports almost all the features, except from the Support Schema feature, which is partially supported by the XQuery language. Since the XQuery language can partially exploit schema information, it supports only path expressions, while can not support type-based queries. Note that the Full Text Search and Update features have been recently introduced as W3C recommendations (see [13] and [12] respectively).

We can observe from Table 4, the SPARQL 1.1 (which is currently a W3C working draft) covers almost all the features which are not supported by the current SPARQL version (SPARQL 1.0). However, the Full Text Search feature is not supported by SPARQL 1.1; several SPARQL implementations (e.g., Jena, Sesame, etc.), though, support full text search. In addition, neither the User-Defined Functions nor the Rich Functions features are supported by the SPARQL 1.1 which particularly supports about 20 build-in functions. Finally, the SPARQL 1.1 does not support the Flexible Result Form and partially supports the Restructure Result features; SPARQL can restructure the results of CONSTRUCT queries, however these results can be only RDF graphs.

**Table 4. Comparison of the XQuery and SPARQL Query Languages**

| Feature | Query Languages | | |
|---|---|---|---|
| | **XQuery** | **SPARQL 1.0** | **SPARQL 1.1** |
| **Paths – Reg. Expr.** | ✓ | ✗ | ✓ [18] |
| **Full Text Search** | ✓ [13] | ✗ | ✗ |
| **Nested Queries** | ✓ | ✗ | ✓ |
| **Aggregate Functions** | ✓ | ✗ | ✓ |
| **Restructure Result** | ✓ | ❖ | ❖ |
| **Flexible Result Form** | ✓ | ✗ | ✗ |
| **Support Schema** | ❖ | ✓ | ✓ |
| **Negation** | ✓ | ✗ | ✓ |
| **User-Defined Functions** | ✓ | ✗ | ✗ |
| **Rich Functions** | ✓ | ✗ | ✗ |
| **Update** | ✓ [12] | ✗ | ✓ [17] |

**Legend:** ✓ Supported ✗ Not Supported ❖ Partially

## 3. Motivating Example — Use Cases

In this section, we outline two "real-word" scenarios in order to illustrate the need for bridging the XML and the SW world. In our examples, three hypothetically autonomous partners are involved: Digital Library X (Fig. 2) which is an audiovisual digital library that belongs to an institution or an organization, as well as two video on demand servers (Fig. 3), VoD Server A and VoD Server B. Each partner provides different content and has adopted different technologies to represent and manage their data.

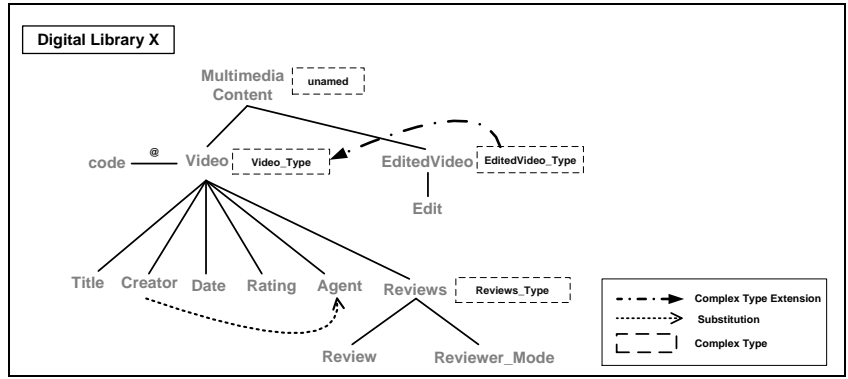

**Fig. 2.** An excerpt of the XML Schema of Digital Library X describing Audiovisual Material

In particular, Digital Library X has adopted XML-related technologies (i.e., XML, XML Schema, and XQuery) and its contents are described in XML syntax, while both servers have chosen SW technologies (i.e., RDF/S, OWL, and SPARQL) for their content. In addition, Digital Library X provides information for audiovisual material from several domains, while the video on demand servers only for specific types of movies (i.e., New Greek Movie and Newly Released and Highly Rated).

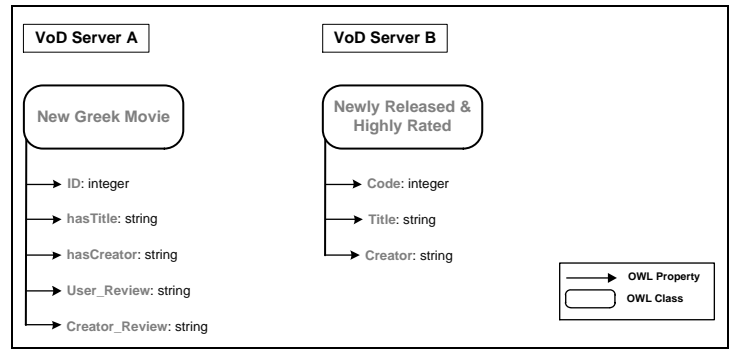

**Fig. 3.** The Ontologies of VoD Server A and VoD Server B describing movies

1st Scenario: *Querying XML data based on an automatically generated ontology.* Consider that Digital Library X wants to publish their data in the Web of Data using SW technologies, a scenario which is very common nowadays. In the Linked Data era, a large number of organizations, institutes and companies (e.g., *DBpedia*, *ACM*, *IEEE*, *IBM*, *NASA*, *DBLP*, *BBC*, *NSF*, *Geonames*, *PubMed*, etc.) publish their data utilizing SW standards and technologies. In particular, they offer SPARQL endpoints (i.e., SPARQL-based search services) over their data. In this case, a schema transformation and a query translation mechanism are required. Using the schema transformation mechanism, the XML Schema of Digital Library X will be transformed to an ontology. Then, the query translation mechanism will be used to translate the SPARQL queries posed over the generated ontology to XQuery queries over the XML data.

As a special case of this scenario, consider the case, in which Digital Library X wants to publish their data in the Web of Data, however, unlike the 1st scenario, Digital Library X wants to use existing, well accepted vocabularies (e.g., Friend of a Friend (*FOAF*)[2], Dublin Core (*DC*)[3], etc.).

2nd Scenario: *Querying XML data based on existing ontologies*. Consider Web of Data users and/or applications, who express their queries or have implemented their APIs over the ontologies of VoD Server A and/or VoD Server B using the SPARQL query language. These users and applications should be able to have access to Digital Library X, without being required to adjust their working environment (e.g., query language, schema, API, etc.).

In these cases, a mapping model and a query translation mechanism are required. In such a case, an expert specifies the mappings between the ontologies of VoD Server A and VoD Server B and the XML Schema of Digital Library X. These mappings are, then, exploited by the query translation mechanism in order to translate the SPARQL queries posed over the ontology of VoD Server A and VoD Server B to XQuery queries to be evaluated over the XML data of Digital Library X.

To sum up, in the Linked Data era, publishing legacy data and offering SPARQL endpoints has become a major research challenge. Although, several systems (e.g., *D2R Server* [80], *OpenLink Virtuoso* [82], *TopBraid Composer*[4], etc.) offer SPARQL endpoints[5] over relational data, to best of our knowledge there is no system supporting XML data (except from the SPARQL2XQuery Framework presented here).

## 4. Bridging the Semantic Web and XML worlds — A Survey

In order to overcome the heterogeneity among the information systems, a large number of *data integration* [23] and *data exchange systems* [24] have been proposed in the literature. The *data integration systems* provide mechanisms for querying heterogeneous sources in a uniform way based on a global schema. The *data exchange systems* (also known as *data transformation/translation systems*) restructure the data from the sources according to a global schema. In recent research works, semantics are exploited to bridge the heterogeneity gap among the information systems and provide semantic integration and interoperability [59][60].

In the context of XML, the first research efforts have attempted to provide interoperability and integration between the relational and XML worlds (e.g., [61][62][63][64][65][66][67]). In addition, several approaches focus on data integration and exchange over heterogeneous XML sources (e.g., [68][69][70][71][72][73][74][75][76]).

---

[2] *http://www.foaf-project.org/*

[3] *http://dublincore.org/schemas/rdfs/*

[4] *http://www.topquadrant.com/products/TB_Composer.html*

[5] Virtual SPARQL endpoints (i.e., with no need to transform the relational data to RDF data).

Regarding the XML and SW worlds, numerous approaches for transforming XML Schema to ontologies and/or XML data to RDF and vice versa have been proposed in the literature. Moreover, some other recent approaches combine SW and XML technologies in order to transform XML data to RDF and vice versa.

In the rest of this section we present the existing approaches that deal with the interoperability and integration issues between the XML and SW worlds (Section 4.1). The recent approaches are described and compared in Section 4.2. Finally, a discussion about the drawbacks and the limitations of the current approaches is presented in Section 4.3.

## 4.1.    Existing Approaches — An Overview

In this section, we present the literature related to interoperability and integration issues between the SW and XML worlds. Table 5 and present the proposed systems in terms of the environment characteristics and the supported operations. These systems have been distinguished into *data integration systems* (Table 5) and *data exchange systems* (Table 6) and are presented in a chronological order.

Table 5 provides an overview of the *data integration* systems in terms of the Environment Characteristics and the supported Operations. In particular, the system described in each row is specified in the first column (System), the environment characteristics are shown in columns 2–4 and the supported operations are shown in columns 5–6. The environment characteristics include the Data Models of the underlying data sources ($2^{nd}$ column), the involved Schema Definition Languages ($3^{rd}$ column) and the supported Query Languages ($4^{th}$ column). The operations include the Query Translation operation ($5^{th}$ column) and the Schema Transformation operation ($6^{th}$ column). If a schema transformation operation is supported the value is the operation description; if the method does not support schema transformation, the value is "no".

Table 6 provides an overview of the *data exchange* systems and is structured in a similar way with Table 5. The system described in each row is specified in the first column (System), the Environment Characteristics are shown in columns 2–3 and the supported Operations used are shown in columns 4–5. The environment characteristics include the Data Models of the underlying data sources ($2^{nd}$ column) and the involved Schema Definition Languages ($3^{rd}$ column). The operations include the Schema Transformation operation ($4^{th}$ column), the indication for the Use of an Existing Ontology ($4^{th}$ column) and the Data Transformation mechanism ($5^{th}$ column). If a schema transformation operation is supported, the value of the third column is the operation description; if the method does not support schema transformation, the value is "no". If the value of the fourth column is "yes", the method supports mappings between XML Schemas and existing ontologies and, as a consequence the XML data are transformed according to the mapped ontologies. Finally, if a data transformation mechanism is provided, the fifth column has its description as value and the value "no" if the system does not provide a data transformation mechanism.

It can be observed from Table 5 that the *data integration systems* are older, thus they do not support the currently standard technologies (i.e., XML Schema, OWL, RDF, SPARQL, etc.). Notice also, that, although the *data exchange systems* shown in Table 6 are more recent, they do not support an integration scenario neither they provide query translation mechanism. Instead, they focus on data and schema transformation, exploring how the RDF data can be transformed in XML syntax and/or how the XML Schemas can be expressed as ontologies and vice versa. In the next section, we describe the latest efforts; most of them focus on combining the XML and the SW technologies in order to transform the underlying data.

**Table 5. Overview of the Data Integration Systems in the SW and XML Worlds**

| Data Integration Systems | | | | | |
|---|---|---|---|---|---|
| **System** | **Environment Characteristics** | | | **Operations** | |
| | **Data Model** | **Schema Definition Language** | **Query Language** | **Query Translation** | **Schema Transformation** |
| STYX (2002) [32][33] | XML | DTD / Graph | OQL / XQuery | OQL → XQuery | no |
| ICS–FORTH SWIM (2003) [34][35][36] | Relational / XML | DTD / Relational / RDF Schema | SQL / XQuery / RQL | RQL → SQL & RQL → XQUERY | no |
| PEPSINT (2004) [37][38][39][40] | XML | XML Schema / RDF Schema | XQuery / RDQL | RDQL → XQuery | XML Schema → RDF Schema |
| Lehti & Fankhauser (2004) [41] | XML | XML Schema / OWL | XQuery / SWQL | SWQL → XQuery | XML Schema → OWL |
| SPARQL2XQuery [95][96] | XML | XML Schema / OWL | XQuery / SPARQL | SPARQL → XQuery | XML Schema → OWL (XS2OWL) |

**Table 6. Overview of the Data Exchange Systems in the SW and XML Worlds**

| Data Exchange Systems | | | | | |
|---|---|---|---|---|---|
| **System** | **Environment Characteristics** | | **Operations** | | |
| | **Data Model** | **Schema Definition Language** | **Schema Transformation** | **Use Existing Ontology** | **Data Transformation** |
| Klein (2002) [42] | XML / RDF | XML Schema / RDF Schema | no | no | XML → RDF |
| WEESA (2004) [43] | XML / RDF | XML Schema / OWL | no | yes | XML → RDF |
| Ferdinand et al. (2004) [44] | XML / RDF | XML Schema / OWL–DL | XML Schema → OWL–DL | no | XML → RDF |
| Garcia & Celma (2005) [45] | XML / RDF | XML Schema / OWL–FULL | XML Schema → OWL–FULL | no | XML → RDF |
| Bohring & Auer (2005) [46] | XML / RDF | XML Schema / OWL–DL | XML Schema → OWL–DL | no | XML → RDF |
| Gloze (2006) [47] | XML / RDF | XML Schema / OWL | no | no | XML ↔ RDF |
| JXML2OWL (2006 & 2008) [48][49] | XML / RDF | XML Schema / OWL | no | yes | XML → RDF |
| XS2OWL (2007) [20][21] | XML / RDF | XML Schema / OWL–DL | XML Schema → OWL–DL | no | XML ↔ RDF |
| GRDDL (2007) [25] | XML / RDF | not specified | no | no | XML ↔ RDF [6] |
| SAWSDL (2007) [26] | XML / RDF | not specified | no | no | XML ↔ RDF [6] |
| Thuy et al. (2007 & 2008) [50][51] | XML / RDF | DTD / OWL–DL | DTD → OWL–DL [6] | no | XML → RDF [6] |
| Droop et al. (2007 & 2008) [28][29][30] | XML / RDF | not specified | no | no | XML → RDF [6] |

---

[6] The transformation is performed in a semi-automatic way that requires user intervention.

**Table 6. (cont'd) Overview of the Data Exchange Systems in the SW and XML Worlds**

| Data Exchange Systems | | | | | |
|---|---|---|---|---|---|
| **System** | **Environment Characteristics** | | **Operations** | | |
| | **Data Model** | **Schema Definition Language** | **Schema Transformation** | **Use Existing Ontology** | **Data Transformation** |
| Janus (2008 & 2011) [52][53] | XML / RDF | XML Schema / OWL–DL | XML Schema → OWL–DL | no | no |
| Deursen et al. (2008) [54] | XML / RDF | XML Schema / OWL | no | yes | XML → RDF [6] |
| XSPARQL (2008) [31] | XML / RDF | not specified | no | no | XML ↔ RDF [6] |
| Cruz & Nicolle (2008) [55] | XML / RDF | XML Schema / OWL | no | yes | XML → RDF |
| XSLT+SPARQL (2008) [56] | XML / RDF | not specified | no | no | RDF → XML |
| DTD2OWL (2009) [57] | XML / RDF | DTD / OWL–DL | DTD → OWL–DL | no | XML → RDF |
| Corby et al. (2009) [58] | XML / RDF / Relational | not specified | no | no | XML → RDF [6] Relational → RDF |
| TopBraid Composer (Maestro Edition) – TopQuadrant [4] (Commercial Product) | XML / RDF | not specified / OWL | XML → OWL | no | XML ↔ RDF [6] |
| XS2OWL 2.0 (2011) [22] | XML / RDF | XML Schema 1.1 / OWL 2 | XML Schema 1.1 → OWL 2 | no | XML ↔ RDF |

## 4.2. Recent Approaches

In this section, we present the latest approaches related to the support of interoperability and integration between the XML and SW worlds. The former approaches utilize the current W3C standards technologies (e.g., XML Schema, RDFS, OWL, XQuery, SPARQL, etc.). With the *DTD2OWL* [57] system be an exception, since it is focus on transforming DTD schemas (instead of XML Schemas) to OWL ontologies. The most of the latest efforts focus on combining the XML and the SW technologies in order to provide an interoperable environment. In particular, they merge SPARQL, XQuery, XPath and XSLT features to transform XML data to RDF and vice versa.

The *Semantic Annotations for WSDL* (*SAWSDL*) W3C Working Group [25] uses XSLT to convert XML data into RDF and a combination of SPARQL and XSLT for the inverse transformation. Additionally, the *Gleaning Resource Descriptions from Dialects of Languages* (*GRDDL*) W3C Working Group [26] uses XSLT to extract RDF data from XML.

*XSPARQL* [31] combines SPARQL and XQuery in order to achieve the transformation from XML to RDF and back. In the XML to RDF scenario, XSPARQL uses a combination of XQuery expressions and SPARQL Construct queries. The XQuery expressions are used to access XML data and the SPARQL Construct queries are used to convert the accessed XML data into RDF. In the RDF to XML scenario, XSPARQL uses a combination of SPARQL and XQuery clauses: The SPARQL clauses are used to access RDF data, and the XQuery clauses are used to format the results in XML syntax.

In *XSLT+SPARQL* [56] the XSLT language is extended in order to embed SPARQL SELECT and ASK queries. The SPARQL queries are evaluated over RDF data and the results are transformed to XML using XSLT expressions.

In other approaches, SPARQL queries are embedded into XQuery and XSLT queries [27]. In [28][29][30], XPath expressions are embedded in SPARQL queries. The former approaches try to process XML and RDF data in parallel and benefit from the combination of the SPARQL, XQuery, XPath and XSLT language characteristics. Also, a method that transforms XML data to RDF and translates XPath queries into SPARQL has been proposed in [28][29][30].

## 4.3. Drawbacks & Limitations — A Discussion

In this section we present a discussion over the existing approaches related to the support of interoperability and integration between the XML and SW worlds and highlight their main drawbacks and limitations.

As already mentioned, the existing data integration systems (Table 5) are quite old and do not support the current standard technologies (e.g., XML Schema, OWL, RDF, SPARQL, etc.). On the other hand, the data exchange systems (Table

6) are more recent. However, they neither support integration scenarios nor they provide query translation mechanisms. Instead, they focus on data transformation. Finally, none of the existing systems can offer SPARQL endpoints (i.e., SPARQL-based search services) over XML data (except from the SPARQL2XQuery Framework presented later in this chapter).

The main drawback of the latest approaches ([25][26][27][28][29][30][31][56]) is that there is no way to express XML retrieval queries using the SPARQL query language.

In particular, the users of these systems are forced to: (a) be familiar with both the XML and SW models and languages; (b) be aware of both ontologies and XML Schemas in order to express their queries; and (c) be aware of the syntax and the semantics of each of the above approaches in order to express their queries. In addition, each of these approaches has adopted its own syntax and semantics by modifying and/or the merging the standard technologies. These modifications may also result in compatibility problems. A consequence of the assumptions made by these approaches is that they have been evaluated over only a few kilobytes of data.

The major limitation of the existing data exchange systems, which provide schema transformation mechanisms, is that they do not support the XML Schema identity constraints (i.e., key, keyref, unique). In addition, none of them supports the XML Schema user-defined simple datatypes. Finally, none of the existing approaches considers the new constructs introduced by XML Schema 1.1. These limitations have been overcome by the Xs2Owl Framework [20][22], which is presented later in this chapter (Section 6).

The Xs2Owl Framework belongs to the data exchange systems category. It provides a transformation model for the automatic and accurate expression of the XML Schema semantics in OWL syntax. In addition, it allows the transformation of XML data in RDF format and vice versa. The current version of the Xs2Owl Transformation Model [22] exploits the OWL 2 semantics, in order to achieve a more accurate representation of some XML Schema constructs. For instance, the XML Schema identity constraints (i.e., key, keyref, unique) can now be accurately represented in OWL 2 syntax (which was not feasible with OWL 1.0), thus overcoming the most important limitation of the previous version of Xs2Owl Transformation Model. In addition, this version supports the new XML constructs introduced by XML Schema 1.1 [4]. To the best of our knowledge, this is the first work that fully captures the XML Schema semantics and support the XML Schema 1.1 constructs.

Regarding the data integration systems and in particular the query translation operation, a major limitation in the existing literature is that there do not exist approaches addressing the SPARQL to XQuery translation.

The SPARQL to SQL translation has been extensively studied and several systems and approaches have been proposed in the literature both for integrating relational data with the SW (e.g., [77][78][79][80][81][82][83]) and for accessing RDF data stored in relational databases (e.g., [84][85][86][87]). However, despite the significant research efforts on the SPARQL to SQL translation, to the best of our knowledge there is no work addressing the SPARQL to XQuery translation.

Given the very high importance of XML and the related standards in the web, this is a major limitation in the existing research.

In the Linked Data era, publishing legacy data and offering SPARQL endpoints has become a major research challenge. Although, several systems (e.g., *D2R Server* [80], *OpenLink Virtuoso* [82], *TopBraid Composer*[4]) offer SPARQL endpoints over relational data, to best of our knowledge there is no system supporting XML data.

The SPARQL2XQuery Framework [95][96] presented in this chapter (Section 5), provides a formal model for the expression of mappings from OWL to XML Schema and a generic method for SPARQL to XQuery translation. The SPARQL2XQuery Framework supports both manual and automatic mapping specification. In addition, it has been integrated with the XS2OWL Framework, thus facilitating the automatic mapping generation and maintenance. Moreover, the SPARQL2XQuery Framework has been evaluated over large datasets.

Compared to the latest approaches ([25][26][27][28][29][30][31][56]), in the SPARQL2XQuery Framework working scenarios the users (a) work only on SW technologies; (b) are not expected to know the underlying XML Schema or even the existence of XML data; and (c) they express their queries only in standard (i.e., without modifications) SPARQL syntax.

## 5. The SPARQL2XQuery Framework — An Overview

In this section, we present an overview of the SPARQL2XQuery Framework [95] [96] that has been developed to provide an interoperable environment between the SW (*OWL/RDF/SPARQL*) and XML (*XML Schema/XML/XQuery*) worlds.

In particular, the SPARQL2XQuery Framework offers an interoperable environment where SPARQL queries are automatically translated to XQuery queries, in order to access XML data across the Web. The SPARQL2XQuery Framework provides a mapping model for the expression of OWL to XML Schema mappings and a method for SPARQL to XQuery translation. To this end, our Framework supports both manual and automatic mapping specification between ontologies and XML Schemas. In order to support the automatic mapping specification scenario, the SPARQL2XQuery has been integrated with our XS2OWL Framework which generates OWL ontologies that fully capture the XML Schema semantics. The system architecture is presented in Fig. 4.

The SPARQL2XQuery Framework provides an essential component for the Linked Data environment that allows setting SPARQL endpoints over existing XML data, as well as a fundamental part of ontology-based integration frameworks involving XML sources.

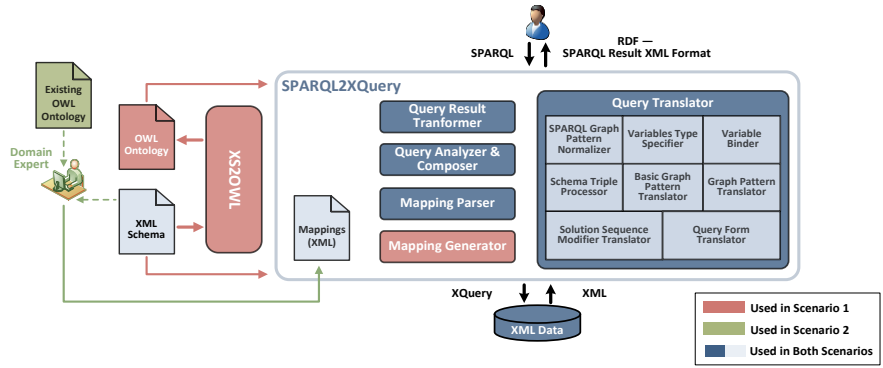

**Fig. 4. System Architecture.** In the first scenario, the XS2OWL Framework is used to generate an OWL ontology from the XML Schema. The mappings are automatically specified and stored. In the second scenario, a domain expert specifies the mappings between an existing ontology and the XML Schema. In both scenarios, SPARQL queries are processed and translated into XQuery queries for accessing the XML data. The results are transformed in the preferred format and returned to the user.

As shown in Fig. 4, our working scenarios involve existing XML data that follow one or more XML Schemas. Moreover, the SPARQL2XQUERY Framework supports two different scenarios:

**1ˢᵗ Scenario: Querying XML data based on automatically generated OWL ontologies.** In that case, the following steps take place:

(a) Using the XS2OWL Framework, the XML Schema is expressed as an OWL ontology.

(b) The Mappings Generator component, taking as input the XML Schema and the generated ontology; automatically generates, maintains and stores the mappings between them in XML format.

(c) The SPARQL queries posed over the generated ontology are translated by the Query Translator component to XQuery expressions.

(d) The query results are transformed by the Query Result Transformer component into the desired format (SPARQL Query Result XML Format [19] or RDF).

In this case, the SPARQL2XQUERY Framework can be utilized as a fundamental component of *hybrid ontology–based integration* [59] frameworks (e.g., [93]), where the schemas of the XML data sources are represented as OWL ontologies and these ontologies are further mapped to a global ontology.

**2ⁿᵈ Scenario: Querying XML data based on existing OWL ontologies.** In this case the following steps take place:

(a) Existing OWL ontologies are manually mapped by a domain expert to the XML Schema.

(b) The SPARQL queries posed over the existing ontology are translated to XQuery expressions.

(c) The query results are transformed into the desired format.

In both scenarios, the systems and the users that pose SPARQL queries over the ontology are not expected to know the underlying XML Schemas or even the existence of XML data. They express their queries only in standard SPARQL, in terms of the ontology that they are aware of, and they are able to retrieve XML data.

Our Framework provides the following operations:

1. **Schema Transformation.** Every XML Schema can be automatically transformed into OWL ontology, using the XS2OWL Framework.

2. **Mappings Generation.** The mappings between the OWL representations of the XML Schemas and the XML Schemas and their OWL representations can be either automatically detected or manually specified and stored as XML documents.

3. **Query Translation.** Every SPARQL query that is posed over the OWL representation of the XML Schemas (first scenario) or over the existing ontology (second scenario), is translated into an XQuery query that can be answered from the XML data.

4. **Query Result Transformation.** The query results are transformed in the preferred format.

5. **XML – RDF Transformation.** Transformation of XML data in RDF syntax and vice versa.

Finally, the SPARQL2XQuery Framework is going to be integrated in an ontology-based mediator framework [91][92][93][94] that we are developing now and is going to provide semantic interoperability and integration support between distributed heterogeneous sources using the standard SW and XML technologies.

## 6. The XS2OWL Framework — An Overview

In this section we describe the schema transformation process exploited in the first scenario supported by the SPARQL2XQuery Framework in order to express the XML Schema semantics in OWL syntax. This is accomplished by the XS2OWL Framework [20][22], which implements the XS2OWL *Transformation Model* . The XS2OWL Transformation Model allows the accurate representation of the XML Schema constructs in OWL syntax without any loss of the XML Schema semantics.

Here we present, an extended and updated version of the XS2OWL Transformation Model, XS2OWL 2.0 [22] exploits the OWL 2 semantics, in order to achieve a more accurate representation of some XML Schema constructs. For instance, the XML Schema identity constraints (i.e., key, keyref, unique) can now be accurately represented in OWL 2 syntax (which was not feasible with OWL 1.0), thus overcoming the most important limitation of the previous version of XS2OWL Transformation Model. Additionally, this version supports the new XML constructs introduced by XML Schema 1.1. To the best of our knowledge, this is the first work that fully captures the XML Schema semantics and support the XML Schema 1.1 constructs.

As shown in Fig. 5, the XS2OWL Framework takes as input an XML Schema *XS* and generates: a) An OWL *Schema ontology* $O_S$ that captures the XML Schema semantics; and (b) A *Backwards Compatibility ontology* $O_{BC}$ which keeps the correspondences between the $O_S$ constructs and the *XS* constructs, and systematically capture the semantics of the XML Schema constructs that cannot be directly captured in $O_S$ (since they cannot be represented by OWL semantics).

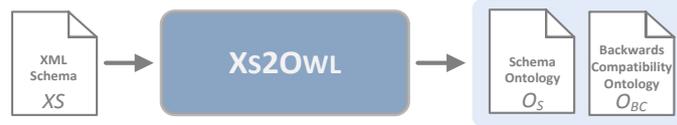

**Fig. 5.** The XS2OWL Framework

The *OWL Schema Ontology* $O_S$, which directly captures the XML Schema semantics, is exploited in the first scenario supported by the SPARQL2XQUERY Framework. In particular, $O_S$ is utilized by the users while forming the SPARQL queries. In addition, the SPARQL2XQUERY Framework processes $O_S$ and *XS* and generates a list of mappings between the constructs of $O_S$ and *XS* (details in Section 7).

**Table 7. Correspondences between the XML Schema Constructs in OWL Syntax, according to the XS2OWL Transformation Model**

| XML Schema Construct | OWL Construct |
|---|---|
| Complex Type | Class |
| Simple Datatype | Datatype Definition |
| Element | (Datatype or Object) Property |
| Attribute | Datatype Property |
| Sequence | Unnamed Class – Intersection |
| Choice | Unnamed Class – Union |
| Annotation | Comment |
| Extension, Restriction | subClassOf axiom |
| Unique (Identity Constraint) | HasKey axiom [*] |
| Key (Identity Constraint) | HasKey axiom – ExactCardinality axiom [*] |
| Keyref (Identity Constraint) | In the Backwards Compatibility Ontology |
| Substitution Group | SubPropertyOf axioms |
| Alternative [+] | In the Backwards Compatibility Ontology |
| Assert [+] | In the Backwards Compatibility Ontology |
| Override, Redefine [+] | In the Backwards Compatibility Ontology |
| Error [+] | Datatype |

**Note.** The [+] indicates the new XML Schema constructs introduced by the XML Schema 1.1 specification. The [*] indicates the OWL 2 constructs.

The ontological infrastructure generated by the XS2OWL Framework, additionally supports the transformation of XML data into RDF format and vice versa [21]. For transforming XML data to RDF, $O_S$ can be exploited to transform XML documents structured according to $XS$ into RDF descriptions structured according to $O_S$. However, for the inverse process (i.e., transforming RDF documents to XML) both $O_S$ and $O_{BC}$ should be used, since the XML Schema semantics that cannot be captured in $O_S$ are required. For example, the accurate order of the XML sequence elements should be preserved; but this information cannot be captured in $O_S$.

A complete listing of the correspondences between the XML Schema constructs and the OWL constructs, as they are specified in the XS2OWL Transformation Model, is presented in Table 7.

## 6.1. *XML Schema Transformation Example*

In this section we present a concrete example that demonstrates the expression of the semantics of the XML Schema in OWL syntax using the XS2OWL Framework. Fig. 6 presents the XML Schema document that corresponds to the XML Schema tree representation of Fig. 2.

```xml
<xs:schema xmlns:xs="http://www.w3.org/2001/XMLSchema">
    <xs:complexType name="Video_Type">
        <xs:group ref="videoGroup"/>
        <xs:attribute name="code" type="xs:integer"/>
    </xs:complexType>

    <xs:complexType name="EditedVideo_Type">
        <xs:complexContent>
            <xs:extension base="Video_Type">
                <xs:sequence>
                    <xs:element name="Edit" type="xs:string" minOccurs="0" maxOccurs="unbounded"/>
                </xs:sequence>
            </xs:extension>
        </xs:complexContent>
    </xs:complexType>

    <xs:complexType name="Reviews_Type">
        <xs:sequence>
            <xs:element name="Review" type="xs:string"/>
            <xs:element name="Reviewer_Mode" type="xs:string"/>
        </xs:sequence>
    </xs:complexType>

    <xs:element name="MultimediaContent">
        <xs:complexType>
            <xs:sequence>
                <xs:element name="Video" type="Video_Type" minOccurs="0" maxOccurs="unbounded"/>
                <xs:element name="EditedVideo" type="EditedVideo_Type" minOccurs="0" maxOccurs="unbounded"/>
            </xs:sequence>
        </xs:complexType>
    </xs:element>

    <xs:group name="videoGroup">
        <xs:sequence>
            <xs:element name="Title" type="xs:string"/>
            <xs:element ref="Creator" minOccurs="1" maxOccurs="unbounded"/>
            <xs:element name="Date" type="xs:date" />
            <xs:element name="Rating" type="xs:float" />
            <xs:element name="Reviews" type="Reviews_Type" minOccurs="0" maxOccurs="unbounded"/>
        </xs:sequence>
    </xs:group>

    <xs:element name=" Creator" type="xs:string"/>
    <xs:element name="Agent" substitutionGroup="Creator" type="xs:string"/>
</xs:schema>
```

**Fig. 6.** XML Schema Document describing Audiovisual Material

The XML Schema of Fig. 6 has the root element MultimediaContent, which is a list of videos and edited videos that has been implemented as an XML Schema sequence that may contain any number of Video elements of type Video_Type and any number of EditedVideo elements of type EditedVideo_Type. The complex type Video_Type represents videos and contains the model group videoGroup and the code attribute, of integer type, which represents the video identification code.

The videoGroup model group is a reusable sequence of elements, including: (a) The Title element, of string type, which represents the title of a video; (b) The referenced top-level Creator element, of string type, which occurs at least once and represents the creator of a video; (c) The Date element, of type xs:date, which represents the date of the video creation; (d) The Rating element, of float type, which represents the average rating score of a video; and (e) The Reviews element, of the Reviews_Type type, which represents the reviews of a video.

The complex type EditedVideo_Type extends the complex type Video_Type and represents edited videos. In addition to the elements and attributes defined in the context of Video_Type, the complex type EditedVideo_Type has the Edit element, of string type, which specifies zero or more edits (e.g., cuts, filters, etc.) that have been applied on the edited video.

The complex type Reviews_Type is a list of reviews, implemented as an XML Schema sequence that may contain any number of Review and Reviewer_Mode elements of string type. These elements represent the review text and the type of the reviewer (e.g., simple user, creator, etc.).

Finally, the top-level element Agent is an element that may substitute the Creator element, as is specified in its substitutionGroup attribute.

Using the XS2OWL Framework, the XML Schema of Fig. 6 is expressed in OWL syntax. The constructs and the semantics of the *Schema ontology* are presented in Table 8 and Table 9. In particular:

- Information about the classes is provided in Table 8, including the class rdf:ID, the name of the corresponding XML Schema complex type and its superclass(es) (rdfs:subClassOf) name(s).
- Information about the datatype properties (*DTP*) and the object properties (*OP*) is provided in Table 9, including the property rdf:ID, the name of the corresponding XML Schema element or attribute, the property type (*DTP* for the datatype properties or *OP* for the object properties), the property domain(s) (rdfs:domain) and the property range(s) (rdfs:range).

**Table 8. Representation of the XML Schema Complex Types in the Schema Ontology ($O_S$)**

| XML Schema Complex Types | Ontology Classes | |
|---|---|---|
| | **rdf:ID** | **rdfs:subClassOf** |
| Video_Type | Video_Type | owl:Thing |
| EditedVideo_Type | EditedVideo_Type | Video_Type |
| Reviews_Type | Reviews_Type | owl:Thing |
| MultimediaContent (*unnamed complex type*) | NS_ MultimediaContent_UNType | owl:Thing |

**Table 9. Representation of the XML Schema Elements and Attributes in the Schema Ontology ($O_S$)**

| XML Schema Elements & Attributes | Ontology Properties | | | |
|---|---|---|---|---|
| | **Type** | **rdf:ID** | **rdfs:domain** | **rdfs:range** |
| Title | DTP | Title_videoGroup__xs_string | Video_Type | xs:string |
| Creator | DTP | Creator__xs_string | Video_Type | xs:string |
| Date | DTP | Date_videoGroup__xs_date | Video_Type | xs:date |
| Rating | DTP | Rating_videoGroup__xs_float | Video_Type | xs:float |
| Agent | DTP | Agent__xs_string | Video_Type | xs:string |
| code | DTP | code__xs_integer | Video_Type | xs:integer |
| Reviews | OP | Reviews_videoGroup__Reviews_Type | Video_Type | Reviews_Type |
| Review | DTP | Review__xs_string | Reviews_Type | xs:string |
| Reviewer_Mode | DTP | Reviewer_Mode__xs_string | Reviews_Type | xs:string |
| Edit | DTP | Edit__xs_string | EditedVideo_Type | xs:string |
| Video | OP | Video_Video_Type | NS_ MultimediaContent _UNType | Video_Type |
| EditedVideo | OP | EditedVideo__EditedVideo_Type | NS_ MultimediaContent _UNType | EditedVideo_Type |
| MultimediaContent | OP | MultimediaContent__NS_ MultimediaContent_UNType | owl:Thing | NS_ MultimediaContent_ UNType |

In addition, the constructs of the Backwards Compatibility ontology generated by the Xs2Owl for the XML Schema of Fig. 6 are presented in Table 10. Notice that the XML Schema constructs are represented in the Backwards Compatibility ontology as individuals of the Backwards Compatibility ontology classes. For every XML Schema construct Table 10 presents: (a) the Class of the corresponding individual in $O_{BC}$; (b) the unique rdf:ID of the individual (3rd column); and (c) the rdf:ID of the corresponding OWL construct in the Schema ontology (4th column).

**Table 10. Representation of the Persons XML Schema Constructs in the Backwards Compatibility Ontology ($O_{BC}$)**

| XML Schema Constructs | Backwards Compatibility Individuals | | |
|---|---|---|---|
| | **Class** | **rdf:ID** | **Schema Ontology Construct rdf:ID** |
| Title | ElementInfoType | Title_videoGroup__xs_string__ei | Title_videoGroup__xs_string |
| Title | DatatypePropertyInfoType | Title_videoGroup__xs_string | Title_videoGroup__xs_string |
| Date | ElementInfoType | Date_videoGroup__xs_date__ei | Date_videoGroup__xs_date |
| Date | DatatypePropertyInfoType | Date_videoGroup__xs_date | Date_videoGroup__xs_date |
| Rating | ElementInfoType | Rating_videoGroup__xs_float__ei | Rating_videoGroup__xs_float |
| Rating | DatatypePropertyInfoType | Rating_videoGroup__xs_float | Rating_videoGroup__xs_float |
| Reviews | ElementInfoType | Reviews_videoGroup__Reviews_Type__ei | Reviews_videoGroup__Reviews_Type |
| Creator | ElementInfoType | Creator__xs_string__ei | Creator__xs_string |
| Creator | DatatypePropertyInfoType | Creator__xs_string | Creator__xs_string |
| Agent | ElementInfoType | Agent__xs_string__ei | Agent__xs_string |
| Agent | DatatypePropertyInfoType | Agent__xs_string | Agent__xs_string |
| Review | ElementInfoType | Reviews_Type_Review__xs_string__ei | Review__xs_string |
| Review | DatatypePropertyInfoType | Reviews_Type_Review__xs_string | Review__xs_string |
| Reviewer_Mode | ElementInfoType | Reviews_Type_Reviewer_Mode__xs_string __ei | Reviewer_Mode__xs_string |
| Reviewer_Mode | DatatypePropertyInfoType | Reviews_Type_Reviewer_Mode__xs_string | Reviewer_Mode__xs_string |
| MultimediaContent | ElementInfoType | MultimediaContent__NS_MultimediaContent_UNType__ei | MultimediaContent__NS_MultimediaContent_UNType |
| MultimediaContent | DatatypePropertyInfoType | MultimediaContent__NS_MultimediaContent_UNType | MultimediaContent__NS_MultimediaContent_UNType |
| Code | DatatypePropertyInfoType | Video_Type_code__xs_integer | code__xs_integer |
| Edit | ElementInfoType | EditedVideo_Type_Edit__xs_string__ei | Edit__xs_string |
| Edit | DatatypePropertyInfoType | EditedVideo_Type_Edit__xs_string | Edit__xs_string |
| Reviews_Type | ComplexTypeInfoType | Reviews_Type | Reviews_Type |
| EditedVideo_Type | ComplexTypeInfoType | EditedVideo_Type | EditedVideo_Type |
| Video_Type | ComplexTypeInfoType | Video_Type | Video_Type |

The XML Schema of Fig. 6 and the schema ontology $O_S$ generated by XS2OWL are depicted in Fig. 7. The correspondences between the XML Schema and the generated ontology are represented with dashed grey lines.

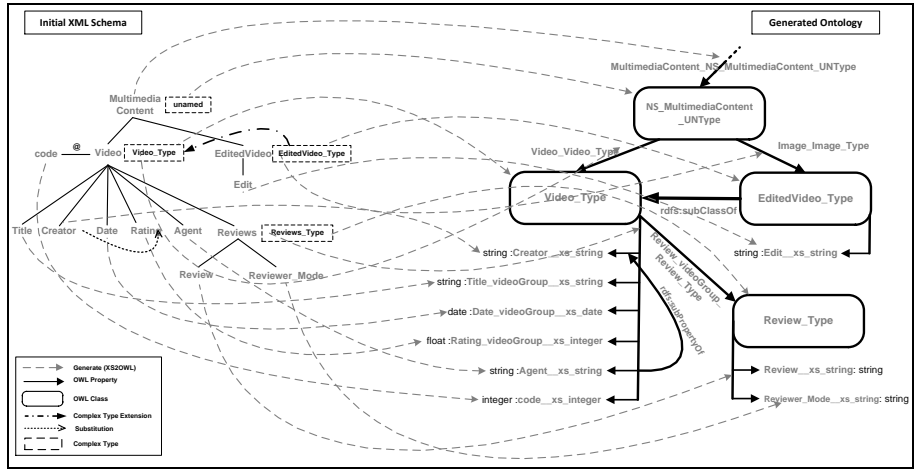

**Fig. 7.** The XML Schema of Fig. 6 and the Schema Ontology $O_S$ generated by XS2OWL with their correspondences drawn in dashed grey lines

## 7. The SPARQL2XQuery Framework — Mapping Model

In this section we outline the mapping model adopted by the SPARQL2XQuery Framework in the context of SPARQL to XQuery translation, in order to allow the expression of mappings between OWL ontologies and XML Schemas.

At the *Schema level* (OWL Ontology/XML Schema), associations between the ontology constructs (i.e., classes, properties, etc.) and the XML Schema constructs (i.e., elements, complex types, etc.) are obtained. Moreover, at the *Data level*, the XML data follow the XML Schema. As a result, the XML Schema construct occurrences in the XML data can be identified and addressed using a set of XPath expressions (*XPath Set*). Thus, based on the correspondences between the ontology and the XML Schema, the ontology constructs are associated with the corresponding XPath expressions (that point to the corresponding XML nodes at the XML data level). Consequently, a *mapping μ* in the context of the SPARQL to XQuery translation is the association of an OWL construct with *XPath Sets*.

In the first scenario supported by the SPARQL2XQuery Framework the mappings are automatically generated. In this case, the XS2OWL Framework is exploited for expressing the semantics of an XML Schema *XS* in OWL syntax, then the mappings between the automatically generated OWL ontology $O_S$ and *XS* are automatically detected, generated and maintained by the SPARQL2XQuery Framework.

The mapping generation is carried out by the *Mappings Generator* component, which takes as input *XS* and $O_S$. The Mappings Generator component parses the input files and obtains a set *M* of mappings, expressed in XML syntax, between all the constructs of $O_S$ and the XPath sets that address all the corresponding XML nodes in the documents that follow *XS*.

In the second scenario, an XML Schema must be mapped with existing ontologies; in this case, the mappings must be manually specified by an expert. Thus, in some cases the XML Schema and the ontology are partially mapped.

## 7.1. *Mapping Examples*

In this section we provide two mapping examples based on both the scenarios supported by the SPARQL2XQuery Framework. In the first example, we present a manual mapping specification case (2nd Scenario), where the VoD server ontologies (Fig. 3) are manually mapped by an expert to the XML Schema of the Digital Library X (Fig. 2). In the second example, we outline the automatic mapping generation between the XML Schema of the Digital Library X and its corresponding OWL ontology that is automatically generated by the Xs2Owl Framework (Fig. 7) (1st Scenario). In this case, the SPARQL2XQuery Framework, automatically generates, maintains and stores the mappings between the XML Schema of the digital library and the generated ontology in XML format.

### Example 1.     *Manual Mapping Specification*

In Fig. 8, excerpts of the VoD server ontologies have been (partially) mapped to the XML Schema of the Digital Library X (2nd Scenario). The mappings are presented with dashed grey lines.

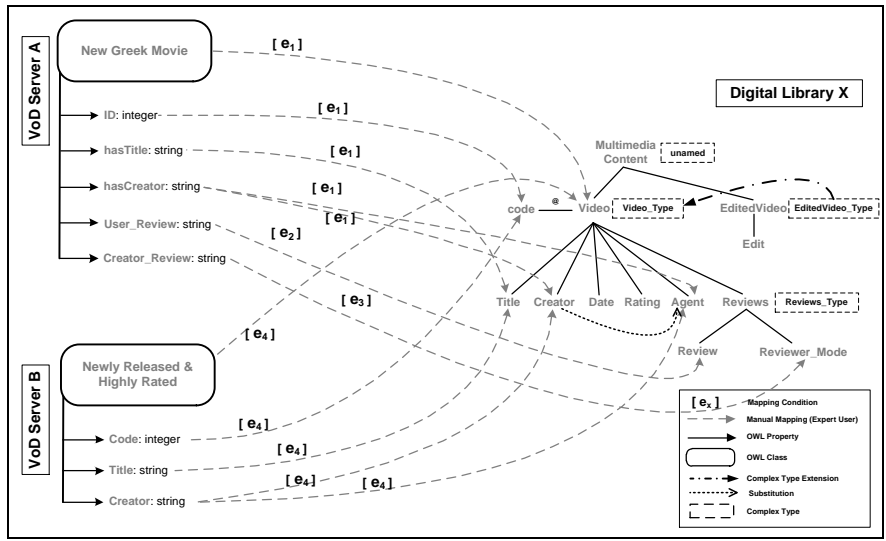

**Fig. 8.** The Existing ontologies of VoD Server A & B are manually mapped with the XML Schema of Digital Library X

It can be observed from Fig. 8 that several mappings have been identified. For instance, the class New Greek Movies from the VoD Server A can be mapped to the XML Schema element Movie, under the condition $e_1$. $e_1$ is a condition that holds for the movies with a `code` attribute that starts with "960" (i.e., Greek code) and a `Date` element equal to "2011". Such a mapping in the context of SPARQL to XQuery translation is of the following form:

μ₁: New Greek Movie ≡ { /*MultimediaContent/ Video[ ./ Date = 2011 and starts-with(. @code, "960") ]* }

In a similar way, the properties `ID` and `hasTitle` can be mapped to the `code` attribute and the `Title` element respectively under the $e_1$ condition. Such mappings are of the following form:

μ₂: ID ≡ { /*MultimediaContent/ Video[ ./ Date= 2011 and starts-with(. @code, "960") ]/ @code* }

μ₃: hasTitle ≡ { /*MultimediaContent/ Video[./ Date= 2011 and starts-with(. @code, "960") ]// Title* }

In addition, the property `User_Review` can be mapped to the `Review` element under the conjunction (i.e., *and*) of the $e_1$ condition with a condition $e_2$ that holds for the reviews with a `Reviewer_Mode` element equal to "User". Also, the `Creator_Review` property can be mapped to the `Review` element under the conjunction of the $e_1$ condition with a condition $e_3$ that holds for the reviews with a `Reviewer_Mode` element equal to "Creator". Such mappings are of the following form:

μ₄: User_Review ≡ { /*MultimediaContent/ Video[ ./ Date =2011 and starts-with(.@code, "960") and ./ Reviews/ Reviewer_Mode = "User" ]// Review* } μ₅: Creator_Review ≡ { /*MultimediaContent/ Video[ ./ Date= 2011 and starts-with(.@code, "960") and ./ Reviews/ Reviewer_Mode = "Creator" ]// Review* }

In the VoD Server B, the class `Newly Released & Highly Rated` can be mapped to the XML Schema element `Video`, under the condition $e_4$. $e_4$ is a condition that holds for videos with a `Date` element equal to "2011" and a `Rating` element greater than "7.5".

μ₅: Newly Released & Highly Rated ≡ { /*MultimediaContent/ Video[ ./ Date = 2011 and./ Rating> 7.5 ]* }

Similarly, the property `Creator` can be mapped to the union of the `Creator` and `Agent` elements under the condition $e_4$.

μ₆: Creator ≡ { /*MultimediaContent/ Video[ ./ Date = 2011 and./ Rating> 7.5 ]// Creator* , /*MultimediaContent/ Video[ ./ Date = 2011 and./ Rating> 7.5 ]// Agent* }

□

***Example 2.*** *Automatic Mapping Generation*

Consider the XML Schema of the `Digital Library X` and the corresponding ontology that has been automatically generated by XS2OWL (Fig. 7). In this case, the *Mappings Generator* component automatically generates the mappings between all the ontology constructs (classes and properties) and the XPath Sets that address the corresponding XML nodes. The generated mappings are listed below.

---

**Generated Mappings**

---

*Classes:*

  $\mu_1$: NS_MultimediaContent_UNType ≡ { */MultimediaContent* }

  $\mu_2$: Video_Type ≡ { */MultimediaContent/Video* }

  $\mu_3$: EditedVideo_Type ≡ { */MultimediaContent/EditedVideo* }

  $\mu_4$: Reviews_Type ≡ { */MultimediaContent/Video/Reviews,  /MultimediaContent/EditedVideo/Reviews* }

*Datatypes Properties:*

  $\mu_5$: code__xs_integer ≡ { */MultimediaContent/Video/@code ,  /MultimediaContent/EditedVideo/@code* }

  $\mu_6$: Creator__xs_string ≡ { */MultimediaContent/Video/Creator ,  /MultimediaContent/EditedVideo/Creator* }

  $\mu_7$: Agent__xs_string ≡ { */MultimediaContent/Video/Agent ,  /MultimediaContent/EditedVideo/Agent* }

  $\mu_8$: Title_videoGroup__xs_string ≡ { */MultimediaContent/Video/Title ,  /MultimediaContent/EditedVideo/Title* }

  $\mu_9$: Date_videoGroup__xs_date ≡ { */MultimediaContent/Video/Date ,  /MultimediaContent/EditedVideo/Date* }

  $\mu_{10}$: Rating_videoGroup__xs_float ≡ { */MultimediaContent/Video/Rating ,  /MultimediaContent/EditedVideo/Rating* }

  $\mu_{11}$: Review__xs_string ≡ { */MultimediaContent/Video/Review ,  /MultimediaContent/EditedVideo/Review* }

  $\mu_{12}$: Reviewer_Mode__xs_string ≡ { */MultimediaContent/Video/Review_Mode,*
  */MultimediaContent/EditedVideo/Review_Mode* }

  $\mu_{13}$: Edit__xs_string ≡ { */MultimediaContent/EditedVideo/Edit* }

*Object Properties:*

  $\mu_{14}$: MultimediaContent__NS_MultimediaContent_UNType ≡ { */MultimediaContent* }

  $\mu_{15}$: Video__Video_Type ≡ { */MultimediaContent/Video* }

  $\mu_{16}$: EditedVideo__EditedVideo_Type ≡ { */MultimediaContent/EditedVideo* }

  $\mu_{17}$: Reviews_videoGroup__Reviews_Type ≡ { */MultimediaContent/Video/Reviews ,*
  */MultimediaContent/EditedVideo/Reviews* }

---

□

# 8. The SPARQL2XQuery Framework — Query Translation

In this section, we present an overview of the SPARQL to XQuery query translation process, which is performed by the *Query Translator* component. The *Query Translator* takes as input a SPARQL query and the mappings between an ontology and an XML Schema and translates the SPARQL query to semantically correspondent XQuery expressions w.r.t. the mappings. To the best of our knowledge, this is the first work addressing this issue.

The query translation process is based on a generic method and a set of algorithms for translating SPARQL queries to XQuery expressions under strict compliance with the SPARQL semantics. The translation covers all the syntax variations of the SPARQL grammar [15]; as a consequence, it can handle every SPARQL query. In addition, the translation process is generic and scenario independent, since the mappings are represented in an abstract formal form as *XPath Sets*. The mappings may be automatically generated (in the first scenario supported by the XS2OWL Framework) or manually specified by a mapping process carried out by an expert (in the second scenario supported by the XS2OWL Framework).

The *Query Translator* component comprises of the following sub-components:

–  The *SPARQL Graph Pattern Normalizer*, that rewrites a Graph Pattern (*GP*) (i.e., the SPARQL `Where` clause) to an equivalent normal form, resulting, in a simpler and more efficient translation process.

– The *Variable Type Specifier*, that identifies the types of the variables, in order to detect any conflict arising from the syntax provided by the user, as well as to identify the form of the results for each variable.

– The *Schema Triple Processor* that processes *Schema Triples* (triples referring to the ontology structure and/or semantics) and binds the appropriate XPaths to the SPARQL variables contained in the Schema Triples.

– The *Variable Binder*, that is used for the assignment of the appropriate XPaths to the SPARQL variables, thus enabling the translation of *GPs* to XQuery expressions.

– The *Basic Graph Pattern Translator*, that performs the translation of *Basic Graph Patterns* (*BGP*) into XQuery expressions.

– The *Graph Pattern Translator*, that translates *GPs* into XQuery expressions. The concept of a *GP* is defined recursively. The *Basic Graph Pattern Translator* sub-component translates the basic components of a *GP* (i.e., *BGPs*) into XQuery expressions, which however have to be properly associated in the context of the *GP* by applying the SPARQL algebra operators (i.e., AND, OPT, UNION and FILTER) among them, using XQuery expressions

– The *Solution Sequence Modifier Translator*, that translates the SPARQL solution sequence modifiers using XQuery clauses and built-in functions.

– The *Query Form Translator*, that is responsible for building the appropriate result structure. SPARQL supports four query forms (i.e., Select, Ask, Construct and Describe). According to the query form, the structure of the final result is different (i.e., an RDF graph, a Result sequence or Boolean value).

## 8.1.   Query Translation Example

Consider a query posed over the ontology of Fig. 7. The query, expressed in both natural language and in SPARQL syntax, is presented below.

---

**Natural Language Query**

---

*"For the instances of the Video_Type class, return the code and the Title, if the creator is "Johnson John", the title includes the string "Music" and the rating is higher than "5". The query must return at most 50 result items ordered by the rating value in descending order and by the id value in ascending order, skipping the first 10 items."*

---

**SPARQL Query**

```
PREFIX ns:   <http://example.com/ns#>
PREFIX rdf:  <http://www.w3.org/1999/02/22-rdf-syntax-ns#>
SELECT ?id  ?title
WHERE {      ?video      rdf:type                        ns:Video_Type .
             ?video   ns:Creator__xs_string           "Johnson John" .
             ?video   ns: code__xs_integer              ?id .
             ?video   ns: Title_videoGroup__xs_string   ?title .
             ?video   ns: Rating_videoGroup__xs_float   ?rating .
             FILTER ( regex( ?title , "Music") && ?rating > 5 )
      } ORDER BY  DESC ( ?rating )  ASC ( ?id )
LIMIT 50  OFFSET 10
```

The SPARQL2XQuery Framework takes as input the mappings of Example 2 and translates the above SPARQL query to the XQuery query presented below.

**Translated XQuery Query**

```
let $doc := collection("http://www.music.tuc.gr/mediaXMLDB/...")
let $Modified_Results :=(
    let $Results :=(
        for $video in    $doc/MultimediaContent/Video[./Creator = "Johnson John" ]
        for $id     in    $video/@code
        for $title  in    $video/Title[matches( . , "Music" )]
        let $rating  :=   $video/Rating[. > 5 ]
        return( <Result> <id>{ string($id) }</id> , <title>{ string($title) }</title> </Result> )
    )
    return ( let $Ordered_Results :=(
                        for $iter in $Results
                        order by $iter/rating descending empty least , $iter/id empty least
                        return($iter) )
            return ( $Ordered_Results[ position( ) > 10 and position( ) <= 60 ]) ) )
)
return ( <Results>{ $Modified_Results }</Results> )
```

## 9. Conclusions

In this chapter we have described the mechanisms that allow the exploitation of the legacy data, in the Web of Data. In particular, in the first part of the chapter (Section 2), we have presented and compared the XML and SW worlds including the technologies and the standards adopted in the two worlds.

In the second part (Sections 3 and 4) we have present a survey of the existing approaches that deal with the interoperability and integration issues between the XML and SW worlds.

Finally, in the third part (Sections 5–8), we have made a brief presentation of the SPARQL2XQuery and XS2OWL Frameworks that have been developed to provide an interoperable environment between the SW and the XML worlds.